# Soluto-thermo-hydrodynamics influenced evaporation of sessile droplets

Abhishek Kaushal [a], Vivek Jaiswal [a], Vishwajeet Mehandia [a] and

Purbarun Dhar [a, b, *]

[a] Department of Mechanical Engineering, Indian Institute of Technology Ropar,

Rupnagar–140001, India

[b] Department of Mechanical Engineering, Indian Institute of Technology Kharagpur,

Kharagpur–721302, India

*Corresponding author:

E–mail: purbarun@mech.iitkgp.ac.in ; purbarun.iit@gmail.com

Tel: +91-1881-24-2119

## Abstract

The present article experimentally and theoretically probes the evaporation kinetics of sessile saline droplets. Observations reveal that presence of solvated ions leads to modulated evaporation kinetics, which is further a function of surface wettability. On hydrophilic surfaces, increasing salt concentration leads to enhanced evaporation rates, whereas on superhydrophobic surfaces, it first enhances and reduces with concentration. Also, the nature and extents of the evaporation regimes (constant contact angle or constant contact radius) are dependent on the salt concentration. The reduced evaporation on superhydrophobic surfaces has been explained based on observed (via microscopy) crystal nucleation behaviour within the droplet. Purely diffusion driven evaporation models are noted to be unable to predict the modulated evaporation rates. Further, the changes in the surface tension and static contact angles due to solvated salts also cannot explain the improved evaporation behaviour. Internal advection is observed (using PIV) to be generated within the droplet and is dependent on the salt concentration. The advection dynamics has been used to explain and quantify the improved evaporation behaviour by appealing to the concept of interfacial shear modified Stefan flows around the evaporating droplet. The analysis leads to accurate predictions of the evaporation rates. Further, another scaling analysis has been proposed to show that the thermal and solutal Marangoni advection within the system leads to the advection behaviour. The analysis also shows that the dominant mode is the solutal advection and the theory predicts the internal circulation velocities with good accuracy. The findings may be of importance to microfluidic thermal and species transport systems.

# 1. Introduction

Understanding the thermo-hydrodynamics and species transport behaviour in microliter droplets has become a focused area of research in recent times. Transport behaviour in microscale droplets has important implications in inkjet printing [1], spray cooling [2], droplet based microfluidic diagnostic tools [3], spray painting [4], microelectronics cooling [5], etc. Droplet evaporation is also important for bio-medical applications like inhalers and nebulizers [6], patterning and detection of ailments from blood [7], DNA/RNA microarrays and nanotechnology [8], etc.

Evaporation kinetics of sessile droplets has been of research interest in fundamental and applied sciences over the last decades. The problem is intriguing due to coupling of heat and mass transfer between liquid and vapour, the associated hydrodynamics and the role of the wetting regimes. The pioneering study on sessile droplet evaporation was by Picknett and Bexon [9]. Two major modes of evaporation, the constant contact radius (CCR) and constant contact angle (CCA) modes were reported. The rate of evaporation was noted to be dependent on the contact radius and contact angle (essentially the wetting state). A vapour diffusion based theory to predict the evaporation rate was also reported.

Bourges and Shanahan [10] discussed the influencing role of droplet evaporation on its transient contact angle (in the CCR mode). The effects of ambient pressure and gas on the evaporation rate have also been studied [11]. Deegan et al. [12] discussed that the evaporating flux is maximum near the contact line, and non-uniformity of the evaporative flux and pinned contact line results in the 'coffee-ring' effect. Popov [13] proposed a closed-form solution for the evaporation rate over the entire range of contact angles. Droplet evaporation on textured substrates has also been explored widely [14]. On such surfaces, the droplet may assume either the Cassie-Baxter state or the Wenzel state of wetting.

In the Cassie-Baxter state [15], the gas phase is trapped below the droplet between pillared structures, thus enhancing the hydrophobicity. In the Wenzel model [16], the surface roughness increases the hydrophobicity by modulating the surface area of contact. McHale et al. [17] and Dash et al. [18] found that droplets superhydrophobic substrate (SHS) also exhibit the three modes of evaporation (CCR, CCA and mixed mode). It was shown that high initial contact angles were not required to ensure CCA mode evaporation [19], as opposed to initial reports. Reports on the influential role of thermal conductivity of the substrate on the evaporation rate of pinned sessile droplets also exist in literature [20].

The variation in internal thermo-hydrodynamics also plays a governing role in altering the evaporation rate of single component droplets. Due to evaporation from the liquid-gas interface, a temperature gradient is established along the interface which leads to thermal Marangoni flows inside the droplet. Hu and Larson [21] reported an analytical model for the non –uniform evaporative flux along the droplet interface which causes the Marangoni stress and consequently the internal advection. Experimental works have [22] described the thermal Marangoni flows within water droplets placed on heated substrates. Fischer [23] found that the enhanced evaporation near the edge of the droplet causes internal flow towards the contact line, thus favouring the coffee ring deposition. Centre enhanced evaporation drives the flow towards the centre, hence supressing coffee ring patterns. In recent times, studies have also explored the evaporation behaviour and the parameters affecting the same for binary mixture of ethanol and water [24-26].

The wetting properties of the substrate also play important roles in the evaporation dynamics of sessile droplets [27]. Droplets on SHS with temperature gradients exhibit two internal counter vortices [28]. Surfactants and colloids are getting more exposure in recent times to understand the role of interfacial hydrodynamics in evaporation [29]. Kang et al. [30] discussed that the internal circulation in drying NaCl droplets on hydrophobic surface is due to the buoyancy driven Rayleigh convection. Karapetsas et al. [31] reported a parametric study to investigate the nature in which the presence of surfactants affects the evaporation process, and the internal hydrodynamics with and without the presence of particles. Hu and Larson [32] observed that the internal circulations due to Marangoni convection supresses particle deposition patterns. The effects of solutal gradients on droplet interfacial dynamics and evaporation kinetics have received attention [33-36]. However, a detailed quantitative approach towards understanding the nature of internal advection, the genesis of the same and the role of wetting states on the evaporation kinetics still remain elusive.

The present work presents detailed experimental and analytical study of evaporation of saline droplets on hydrophilic and SHS. The complex physics behind the multi-component system has been clearly segregated and the main mechanisms have been identified. The roles of salt concentration and solubility on the evaporation kinetics have been brought out. Further, the alteration in the internal advection due to presence of salt has been diagnosed using Particle Image Velocimetry (PIV). A mathematical formulism using scaling analysis has been proposed to model the observed kinetics. The solutal Marangoni advection is noted to be dominant over the thermal counterpart and agrees well to PIV observations. Since vapour-diffusion based model is inadequate to explain the modified evaporation rate, a simplistic Stefan flow based model has been proposed, and noted to agree well with experiments.

## 2. Materials and methodologies

A customized experimental setup (refer fig. 1) is used to study the evaporation kinetics process of sessile droplets on surfaces with different wettability. Sodium Iodide (NaI) and Copper Sulphate pentahydrate ($CuSO_4.5H_2O$) (procured from Merck, India) solutions (in DI water) of concentrations 0.005, 0.01, 0.1 and 0.25 M is used. The salts are selected based on previous reports by the present authors [37]. Cleaned, sterile glass slides are used as hydrophilic surfaces (contact angle for water ~ 40°) and the SHS (contact angle for water ~ 155°, and roll off angle ~3–4°, and very minor contact angle hysteresis ~ 3–5°) is synthesized by spray coating (Rust Oleum Industrial brands, USA) glass slides. The value of solid-liquid interfacial energy $\sigma_{sl}$ is obtained using Young's equation and assuming the same surface energy value of solid-gas interactions ($\sigma_{sg} = 0.375$ J/m$^2$) for glass [38] and the SHS substrate.

**Table 1:** The liquid-gas and solid-gas components of surface energy and static contact angles for the substrates

| Substrate | $\sigma_{sl}$ (J/m$^2$) | $\sigma_{lg}$ (J/m$^2$) | Static Contact angle |
|-----------|-------------------------|-------------------------|----------------------|
| Glass     | 0.316                   | 0.0728                  | 40° ± 3°             |
| SHS       | 0.441                   | 0.0728                  | 155° ± 3°            |

A digitized precision droplet dispensing mechanism (Holmarc Opto-mechatronics, India) has been used. The droplet is dispensed carefully on the substrate from a micro-litre glass syringe (capacity 50 ± 0.1 µL) attached to the dispensing mechanism. The volume of the droplet used in experiments is 20 ± 0.5 µL. This ensures that the contact diameter of sessile droplets is equal to or less than the capillary length scale for water. The evaporation process is recorded using a monochromatic CCD camera (Holmarc Opto-mechatronics, India) with long distance microscope lens. The camera is mounted on three-axis translation stage capable of 30 fps recording at 1 megapixel resolution.

A brightness controlled LED array (DPLED, China) is used for backlight illumination. The evaporation process is recorded at 1280 x 960 pixels at 10 fps. The frames are processed using ImageJ (open source software) using macro subroutines to obtain geometric parameters. Spherical cap assumption is used to determine instantaneous droplet volume, contact radius and contact angle. The complete experimental setup is lodged inside an acrylic chamber and placed on a vibration free table to suppress all ambient disturbances. A digitized thermometer and hygrometer is used to note the temperature and humidity conditions 10 mm away from the droplet (using a sensing probe). For all experiments, the temperature varied as 25 ± 2 ºC and the relative humidity varied as 50 ± 5%.

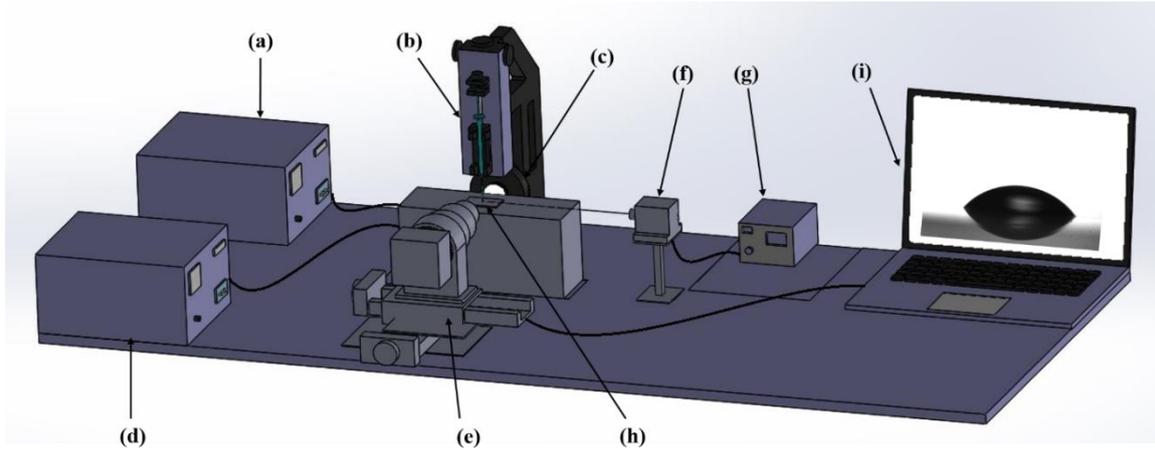

**Figure 1**: Schematic of the experimental setup (a) droplet dispensing mechanism controller unit, (b) droplet dispensing mechanism, (c) LED backlight assembly, (d) backlight illumination controller, (e) CCD camera with long distance microscope lens and three-axis movement control, (f) laser with light sheet optics assembly (not illustrated), (g) laser controller (h) substrate with droplet, (i) computer for data acquisition and camera control. The setup is enclosed in an acrylic chamber.

To diagnose the internal flow behavior during the evaporation process, PIV was used. Non-reactive, neutrally buoyant, fluorescent seeding particles (~10μm diameter, Cospheric LLC, USA) were used. A continuous laser (532nm wavelength, 10 mW power) is used for illumination (Roithner GmbH, Germany). A laser sheet of thickness ~1 mm using a plano-convex lens is employed to observe the droplet mid-plane advection. The thickness of the light sheet is on the higher side and it is possible that estimation errors of 20-25 % are possible in the velocimetry studies. For PIV studies, camera resolution of ~120 pixels/mm and 20 fps was used. The study was done for initial few minutes of the evaporation process so that the change in salt concentration is minimal. For hydrophilic surface, the PIV is done using a fluorescent microscope (attached with a CMOS monochrome camera (Sony Corpn.)) at 30 fps.

Since the side view PIV of the hydrophilic droplets was not proper and of low resolution, the top view micro-PIV methodology was used (at 10 X optical zoom). A fluorescent light was used for illumination. A cross-correlation algorithm, with four pass windows of 64, 32, 16 and 8 pixels has been used in the open source code PIV Lab. A stack of 1000 consecutive images are investigated for maximizing the signal-to-noise ratio and spatially averaged velocity contours are obtained. Standard noise reduction pre-processing algorithms are employed to enhance signal-to-noise ratio and peak locking. Infrared imaging (at 4X thermal lens zoom) has been used to determine the thermal gradients within the evaporating droplets (FLIR T650sc thermal camera). It employs an infrared detector of resolution 640 x 512 pixels and an accuracy count of ±0.02 K.

## 3. Results and discussions

### 3. a. Evaporation kinetics of saline droplets

Figure 2 illustrates the time evolution arrays of the droplet during the evaporation process. The NaI solution droplets evaporate faster compared to water droplets on hydrophilic surfaces (fig. 2 (a)), and the increment in evaporation rate is directly proportional to the salt concentration. Increase in evaporation rate with increasing salt concentration has also been noted in pendent droplets [37, 39-40]. The wetting state of the droplet also changes with addition of salt and has also been noted in literature [41-43]. However, in case of SHS (fig. 2 (b)), a counter-intuitive phenomenon is noted. The rate of evaporation enhances in the dilute regime (up to 0.05 M) and beyond that the evaporation rate is reduced (at 0.25 M the rate is similar to that of water droplet).

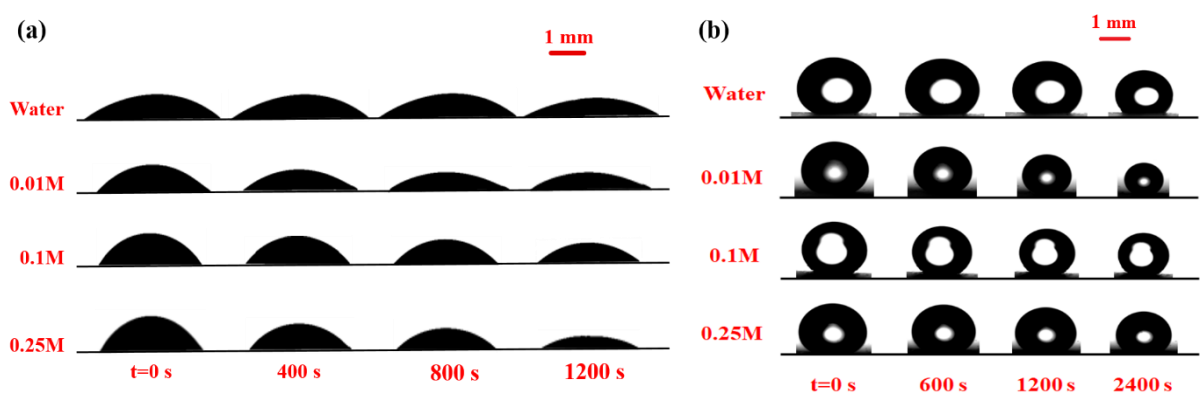

**Figure 2:** Time evolution array of evaporating DI water and NaI solution droplets on (a) hydrophilic surface and (b) SHS. The scale bar represents 1mm for all cases.

The evaporation kinetics can be characterised using the volume reduction law [20] (refer supplementary information for mathematical details), which is expressible as,

$$V^* = \left(\frac{V}{V_0}\right)^{\frac{2}{3}} = 1 - kt \quad (1)$$

Where, V is the instantaneous volume, $V_0$ is the initial volume and k is the evaporation rate constant for a sessile droplet. Figure 3(a) illustrates the temporal evolution the non-dimensional V* for water and saline droplets on hydrophilic substrate. The droplet evaporation time is noted to be significantly decreased and is a direct function of salt concentration. Additionally, the very kinetics of the process has been modulated. The water curve shows three distinct regimes (nearly linear till t=1200 s, non-linear up to ~2300 s, followed by a steep reduction (the rush-hour regime of evaporation)).

The 0.005 M case, on the contrary, exhibits acute example of slip-stick behaviour, with the 2$^{nd}$ regime being characterized by intermittent volume reduction. With increased concentration, only 2 regimes are prominent, and this is due to the reduced wetting state at high concentrations (refer fig. 2 (a)). Figure 3(b) shows the temporal evolution of V* on the SHS. The water droplet shows an initial non-linear regime (up to ~1200 s, and a linear regime thereafter). The dilute (0.005M) solution evaporates faster, however, shows a fully linear regime, while the 0.01M solution exhibits a completely altered non-linear nature. It is noteworthy that eqn. 1 valid within the initial CCR mode of evaporation, and partly in the mixed mode. In the final CCA mode, the curve deviates becomes non-linear, and hence the evaporation rate is no longer constant with time. But, as seen from fig. 2 (a), quasi-linear behaviour may still be modelled (however with different slopes) in the CCA regime for each droplet.

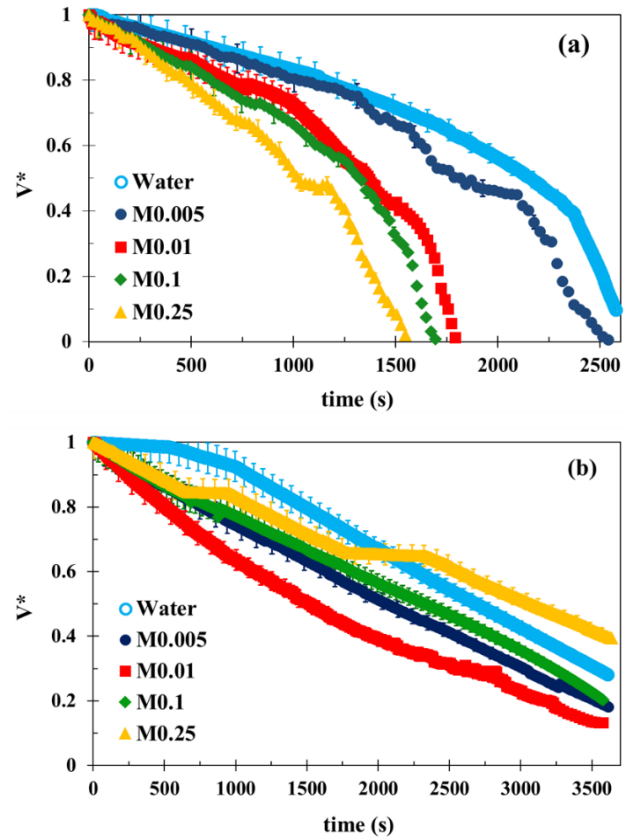

**Figure 3**: Transient variation of V* (refer eqn. 1) for NaI solution droplets with error bar of 5% on (a) hydrophilic surface and (b) SHS.

For sessile droplets, the change in volume does not reflect the exact behaviour of the drop profile. Figure 4(a) illustrates contact angle (normalized with initial contact angle) vs. time on hydrophilic substrate while figure 4(b) shows the same on SHS. It can be observed (refer fig. 4(a)) that increase in the salt concentration increases the rate of reduction of contact angle for both the wetting states. In the hydrophilic case, a continuous reduction in the contact angle is present in the initial stage of evaporation process (CCR mode). Towards the very end, the contact angle curve tends to achieve constant value in the CCA mode. However, this region is largely diminished at higher salt concentrations.

In the case of SHS, the CCA mode is observed during the very initial stages in low concentration droplets. Concentrated droplets show continuous drop in contact angle from the initiation of evaporation and the CCA mode is absent. After the initial regime, the droplet suddenly retracts and tries to gain the original non-wetting shape, which results in intermittent and sudden increase in the contact angle after regular intervals (refer fig. 4(b)). As the concentration of salt increases, the sudden spike in contact angle increase is noted to be frequent and more prominent. Figures 5 (a) and (b) illustrate the behaviour of the non-dimensional contact diameter ($d^* = d/d_0$) of the droplet during evaporation. Here $d$ is the instantaneous contact diameter and $d_0$ is the initial diameter. Two out of three different modes of evaporation (CCR, mixed and CCA) can be clearly observed on hydrophilic surfaces (figure 5(a)) for water and dilute saline droplets.

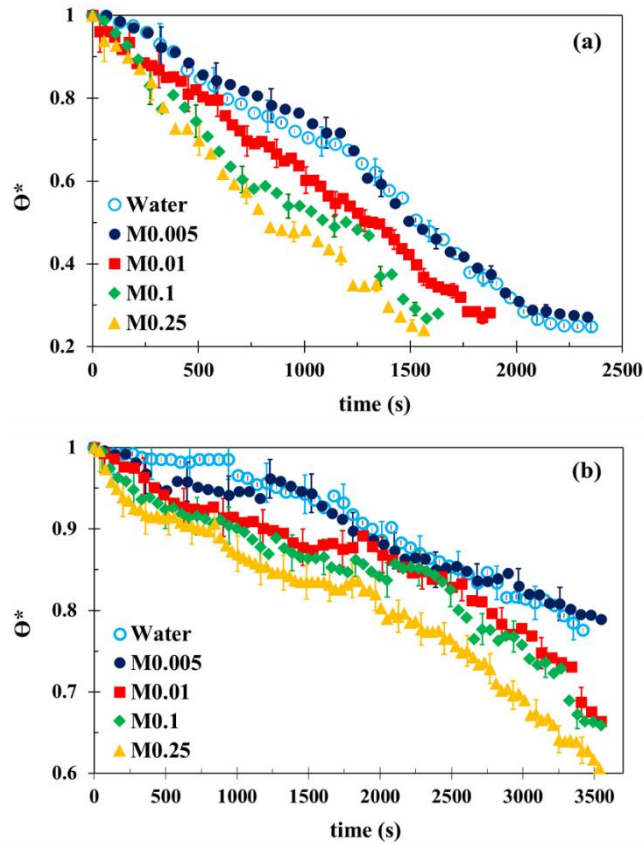

**Figure 4:** Variation of the non-dimensional contact angle with time for NaI solution droplets with error bar of 5% on (a) hydrophilic surface (b) SHS.

Initially the CCR mode is predominant, followed by a short mixed mode regime, in which contact line de-pinning occurs and contact diameter starts decreasing. As the salt concentration increases, the CCR mode regime reduces, which leads to the inference that the propensity of de-pinning of the contact line is enhanced with salt. This could be a direct consequence of the reduced wettability on hydrophilic surfaces with addition of salts (refer fig. 2). On the SHS, the water droplets do not exhibit the well-defined CCR mode. The dilute saline droplets show a complete disappearance of the CCR mode, with enhanced rate of reduction in the droplet diameter. At higher concentrations, the CCR mode appears, but in a piece-wise manner, with intermittent regions of sudden retraction (fig. 5 (b)). Additionally, the rate of reduction of droplet diameter reduces, to the end that the 0.25 M droplet behaves very similar to the water droplet, but with a major presence of the CCR mode.

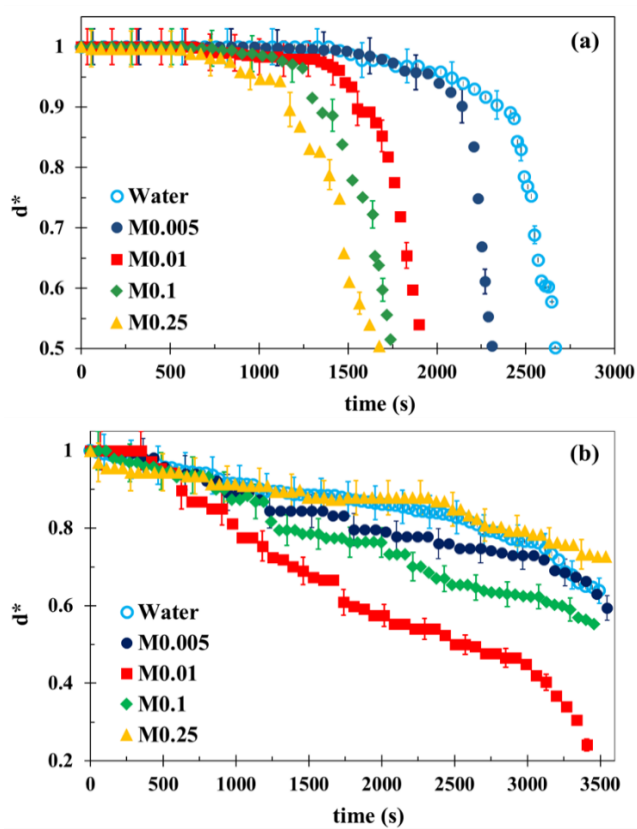

**Figure 5**: Variation of d* with time for NaI solution droplets with error bar of 5% on (a) hydrophilic surface and (b) SHS.

**3. b. Role of interfacial behaviour and diffusion driven evaporation**

The classical diffusion driven models have been used to predict the experimental evaporation rates. The models described by Picknett and Bexon [9] and Shanahan [10] (the former is a purely vapour diffusion based models and the latter is a model based on the geometry of the droplet (refer supplementary information section S1 B)) have been tested. The comparisons with the experimental observations have been illustrated in fig. 6. The models are found to be incapable to predict the augmented evaporation rates, and this establishes that the mechanism at play is not due to changes on the gas side diffusion layer concentration profile. The Picknett and Bexon model can predict the water evaporation rate to good extent, which shows that the saline droplet evaporation enhances and is not an artefact. Hence further probing of the mechanism at play is essential.

    Modulated surface tension and wetting states can also have a direct influence on the evaporation rate of the droplet. The surface tension and contact angles of the saline droplets have been measured using the pendant and sessile drop methods, respectively (refer supplementary information, figs S2 and S3). With addition of salt, the surface tension increases (within 5 %). Improved surface tension might lead the droplet to evaporate faster, as faster reduction in surface area towards a smaller shape is energetically more favourable

from thermodynamics principles. On the contrary, low surface tension fluids (like alcohols) are in general more volatile. Hence, the change in surface tension does not provide any conclusive information regarding the augmented evaporation rate. Also on SHS, the evaporation initially enhances and then reduces, and hence surface tension change is not a major mechanism.

The change in the wetting state (changed contact angle) with salt addition may also be a mechanism behind the enhanced evaporation. The addition of salt is noted to enhance the contact angle (supplementary information, fig S3) for both hydrophilic surface and SHS. While on the hydrophilic surface the evaporation steadily enhances with addition of salt, on the SHS it increases and then decreases with salt addition. Again, droplets on superhydrophobic surfaces are known to evaporate slower (due to reduced surface area), and hence increased contact angles should result in reduced evaporation. Hence, the change in wetting behaviour is also not a very robust mechanism to explain the observations. Thus, the exterior of the droplet and interfacial property modulation are not responsible mechanisms and the interior of the droplet requires probing.

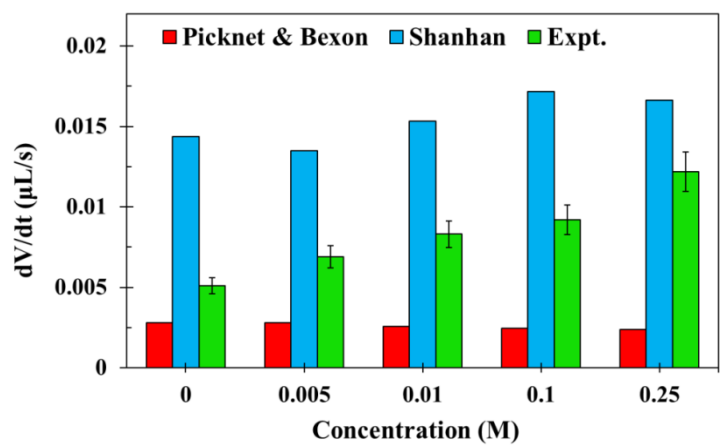

**Figure 6**: Prediction of the evaporation rates from different purely-diffusion-driven models compared with experimental results. The equations used for prediction are described in supporting information (section S1 B).

### 3. c. Influence of internal hydrodynamics

Particle image velocimetry (PIV) studies are performed to quantify the internal behaviour of evaporating saline sessile droplets. The PIV experiments are performed within the initial 5 minutes of initiation of evaporation such that the concentration of salt is not changed largely to induce artefacts (change in concentration is within 10% during PIV studies). For the SHS, the PIV is performed at the vertical mid-plane of the droplet. On hydrophilic surface, the optical clarity and visualization of the vertical mid-plane is difficult due to the wetting posture of the droplet. Here the PIV is done using a microscope arrangement and the horizontal mid-plane of the droplet is studied. Figures 7 (a), (b) and (c) illustrate the velocity contours and vector field for 0.01 M, 0.1 M and 0.2 M saline droplets on SHS and figures 7 (d) and (e) illustrate the same for 0.01 M and 0.2 M droplets on hydrophilic substrate. The PIV is done at 10 fps for 90 seconds and the velocity contours are obtained by temporally averaging the velocity fields for the whole set.

On SHS, water droplets show internal advection (average velocity ~ 0.15 cm/s, not illustrated in figure) with consistent directionality of circulation. It is observed (fig. 7) that the saline droplets (0.01 and 0.1 M) show strong internal circulation, with a distinct advection pattern at the droplet mid-plane. Typically, a large advection cell is present with the circulation axis passing close to the droplet's centre. Interestingly, however, the advection is largely arrested in the 0.2 M case. In the hydrophilic case, the advection in case of water droplet is absent, with mild drift of the seed particles noticeable (not illustrated). With addition of salt, however, detectable and consistent advection is observed (the direction being from the droplet periphery towards the centre, at the plane of visualization). At 0.2 M, the advection pattern becomes more exotic (refer fig. 7 (e)), where a prominent circumferential circulation (represented by green dotted arrow) appears alongside the existing rim to centre advection. The circumferential advection cell cements the notion that Marangoni circulation is appreciable at high concentrations, and the solutal advection (internal circulation) is prominent due to presence of salts.

The internal advection of the droplet (and the interfacial advection on the liquid side) leads to shear at the interface. The shear leads to shearing of the interface on the gas side as well (from balance of shear across the interface). The shear generated within the gas phase leads to advection within the vapour layer shrouding the droplet, which replenishes the layer with ambient air. This improves the species concentration gradient within the otherwise stagnant layer, leading to improves evaporation from the droplet surface [37, 44]. Thus the internal advection is the mechanism behind the augmented evaporation rate. As the advection is arrested for 0.2 M droplet on SHS, the evaporation rate also reduces simultaneously. The counter-intuitive reduction of advection strength has been determined using microscopy.

The sprayed SHS used in present study has a microstructure characterized by micro-cracks and crevices (fig. S5 (a) in supplementary information). The microscope objective is focussed at the contact region between the droplet and the SHS, through the transparent droplet (aided by illuminated bright-field background below the translucent coated glass slide). It is observed that after some time, the cracks act as nucleation sites for inception of crystal growth. Minute crystals are noted to be formed along the cracked regions (refer fig. S5 (b)), which causes largely suppressed internal advection due to presence of solid pseudo-crystal obstacles at the droplet base.

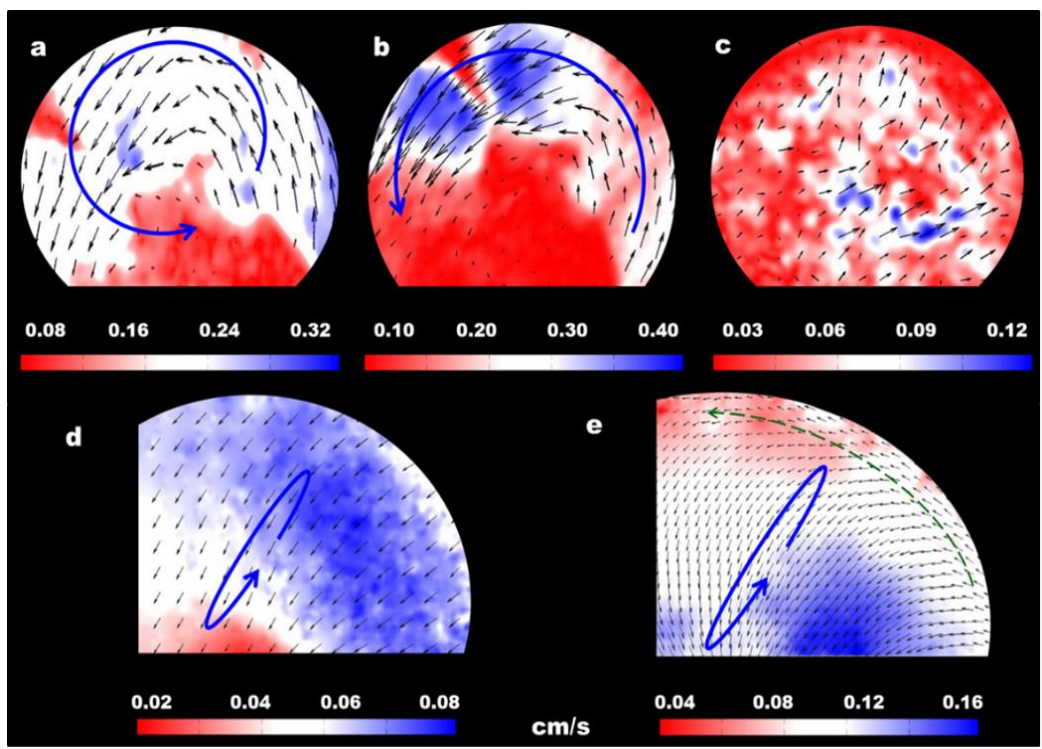

**Figure 7**: Temporally averaged velocity contours and vector fields for saline droplets on SHS (a) 0.01 M (b) 0.1 M (c) 0.2 M, and on hydrophilic surface (top-views) (d) 0.1 M (e) 0.2 M. The large arrows show the nature of the dominant advection currents.

### 3. d. Behaviour of the internal thermal advection

The internal advection is confirmed to be responsible factor towards modulated evaporation kinetics. The genesis of the advection however, remains to be understood. One possibility is thermal advection. Fig. 8 (a) and (b) illustrate the experimental (via infrared imaging) non-dimensional temperature distribution (along non-dimensional contact radius) within the droplets. The distribution corresponds to data within the first five minutes of initiation of evaporation (to overlap closely with the PIV time frame). The associated thermal images are provided as Fig. S4 (supporting information), where clear modulation in the thermal distribution is notable for saline droplets compared to water droplets. For all concentrations, the centre of the droplet is coldest and the temperature increases towards the droplet-vapour interface, and the distribution is due to evaporative cooling of the droplet's bulk.

The water droplet shows a nearly linear behaviour of the thermal profile; while considerable drop in temperature is noted for the 0.1 and 0.2 M droplets (fig. 8 (a)) towards the droplet centre. However, towards the rim, the thermal distribution is similar to water case. This leads to a largely non-linear thermal distribution compared to water. On the contrary, for the SHS, the difference in the thermal profile is similar towards the centre, and different towards the droplet-gas interface. The difference could be attributed to the stronger internal advection (and thus mixing) within droplets on SHS compared to the hydrophilic cases (fig. 7). The role of advection is supported by the fact that the thermal gradient in 0.2 M droplet on SHS behaves similar to the water case, and PIV shows that both have similar internal advection behaviour. The infrared imaging thus establishes that thermal gradients exist within the droplet, and thermal Marangoni convection could be important.

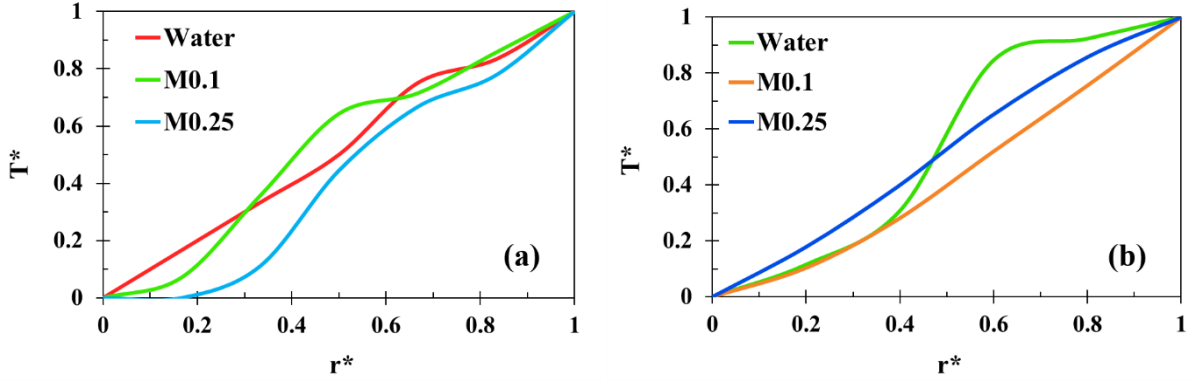

**Figure 8**: Behaviour of the non-dimensional temperature across the droplet (in terms of non-dimensional contact radius) for water and saline solutions (obtained via infrared imaging) on (a) hydrophilic surface and (b) SHS. Here $T^* = \frac{(T-T_{min})}{(T_{max}-T_{min})}$, where min and max represent the minimum and maximum temperatures within the droplet.

A mathematical formulation based on the scaling of different mechanisms of internal thermal advection due to evaporation has been proposed. The energy transferred off an evaporating sessile droplet (contact diameter $d_c$ and contact angle $\theta$) at any instant (LHS of eqn. 2) can be balanced by three thermal transport components within the droplet, viz. Energy transport due to heat diffusion, due to thermal advection (internal flow) [45, 46] and change of surface energy at the liquid-solid interface (RHS components of eqn. 2), and can be expressed as,

$$\dot{m} h_{fg} = 2k A_c \Delta T_M \frac{\cot(\theta/2)}{d_c} + \rho C_p u_M \Delta T_M A_c \sec^2(\theta/2) - A_c \sigma \dot{\theta} \sin\theta \qquad (2)$$

Which can be further expanded into its final form as (refer supplementary information)

$$\rho h_{fg} d_c^2 \dot{\theta} \sec^4(\theta/2) = 16k\Delta T \cot(\theta/2)\left\{Ma \sec^2(\theta/2) - 8Ma\left(\frac{Ja}{Ca}\right)\sin\left(\frac{\theta}{2}\right)\cos\left(\frac{\theta}{2}\right)\right\} \quad (3)$$

Where $\dot{m}$, $h_{fg}$, $k$, $C_p$, $\rho$, $\sigma$, and $u_m$ denote the rate of evaporative mass reduction, the enthalpy of vaporization, the thermal conductivity of the liquid, specific heat of the liquid, density of the liquid, and the average internal circulation velocity, fluid surface tension respectively. $\Delta T_M$ represents the temperature difference caused by evaporative cooling that drives the thermal Marangoni flow due to the surface tension gradient caused by temperature difference. The derivation of the eqn. 2 has been discussed at length in the supporting information (section S2).

The internal circulation velocity is scaled as $u_M = \frac{\sigma_T \Delta T_m}{\mu}$, where $\sigma_T$ is the gradient of surface tension with temperature [31] (value obtained from established correlations), and $\mu$ is the viscosity of the liquid [37, 44]. $\Delta T_M$ can be expressed as (refer supplementary information)

$$\Delta T_M = \left(\frac{\mu h_{fg} d_c \dot{\theta}}{4\sigma_T C_p} sec^2(\theta/2)\right)^{1/2} \text{ for hydrophilic substrate} \qquad (4)$$

$$\Delta T_M = \left(\frac{\mu h_{fg} \dot{R}}{\sigma_T C_p}\right)^{1/2} \text{ for SHS} \qquad (5)$$

Where, $\dot{\theta}$ is the rate of change of contact angle during evaporation. The $\Delta T_M$ for droplets on SHS is determined considering the droplets as spherical systems and that $\dot{R} = \frac{dR}{dt}$. $Ma_T$, $Ca$, and $Ja_e$ represent the thermal Marangoni number, the Capillary number and the evaporation Jacob number, respectively. Mathematically, the numbers can be expressed in terms of droplet parameters as (refer supplementary information)

$$Ma_T = \frac{\sigma_T \Delta T_M d_c}{2\mu\alpha} tan(\theta/2) \qquad (6)$$

$$Ca = \frac{\mu d_c \dot{\theta}}{4\sigma_T} sec^2(\theta/2) \qquad (7)$$

$$Ja_e = \frac{\mu \dot{\theta}}{\rho h_{fg}} \qquad (8)$$

Where $\alpha$ is the thermal diffusivity of the liquid. The Marangoni number governs the thermos-interfacial transport within the droplet, while the Capillary and Jacob numbers illustrate the role of the surface wettability on the behaviour of the evaporative flux. The thermal advection within the droplet can also be caused by the Rayleigh advection, generated due to thermal gradient induced buoyant effects. A similar scaling is performed to determine the related model parameters. The velocities are scaled as buoyancy induced currents driven by the temperature gradient within the droplet. The Rayleigh number (liquid side) can be accordingly scaled as [37, 44],

$$\Delta T_R = \left(\frac{\mu h_{fg} \dot{\theta}}{\rho g \beta d_c C_p} sec^2(\theta/2)\right)^{1/2}, \qquad (9)$$

$$u = \frac{\rho g \beta \Delta T_R d_c^2}{4\mu} \qquad (10)$$

$$Ra = \frac{d_c^2}{8\alpha}\left(\frac{\rho g \beta d_c h_{fg} \dot{\theta}}{\mu C_p} sec^2(\theta/2)\right)^{1/2} \qquad (11)$$

Where g is the acceleration due to gravity, $\beta$ is the coefficient of thermal expansion of the liquid and $\Delta T_R$ is the temperature difference that drives the buoyant advection. Enhanced evaporation due to Rayleigh convection in the ambient gas phase is also possible, however analysis reported [44] shows that the magnitude of this advection is negligibly weak compared to the internal advection.

Hence, both thermal Ma and Ra based advection in the droplet can enhance the evaporation rate. However, the dominant mode requires to be understood. For this, stability analysis proposed by Nield [45] and Davis [46] to quantify thermo-advection in droplets and films has been used. Balance of forces between the two modes is essential to lead to advection within such systems. The analysis is based on the critical Marangoni number ($Ma_c$) and critical Rayleigh number ($Ra_c$). The mathematical requirement for stable advection is

$$\frac{Ma}{Ma_c} + \frac{Ra}{Ra_c} = 1 \qquad (12)$$

The $Ma_c$ for such systems is ~ 80 [44, 45, 46], while the $Ra_c$ is ~ 1708 as per Chandrasekhar's classical analysis. According to Nield and Davis stability analysis, the location of the points on a $Ma_T$ vs Ra plot can provide the regime of advection and the dominant mode behind the same.

Figure 9 illustrates the $Ma_T$ vs Ra maps for different droplets studied. The maps have two regimes given by lines joining $Ra_c$ ~1708 to $Ma_c$ ~80 (as per Nield) and $Ma_c$ ~52 (as per Davis). Points lying below the Davis line (D) represent unstable advection, the ones in between D and Nield (N) represent intermittent, partially stable thermal advection, and those above the N represent stable advection. It is noted that the associated Ra are very less compared to the $Ra_c$, signifying no role of internal thermal Rayleigh advection. On hydrophilic surface, addition of salt leads to shift of the data points towards higher Ma, however, the regime is still in unstable and weak advection. On the SHS, the data points shift further compared to the hydrophilic, however, the regime is still of unstable advection. The thermal Rayleigh advection has already been ruled out as a weak mechanism, and the thermal Marangoni advection also proves to be incapable to induce stable internal advection (which contradicts the PIV observations). Hence, thermal advection is ruled out as the governing mechanism and further probing is essential.

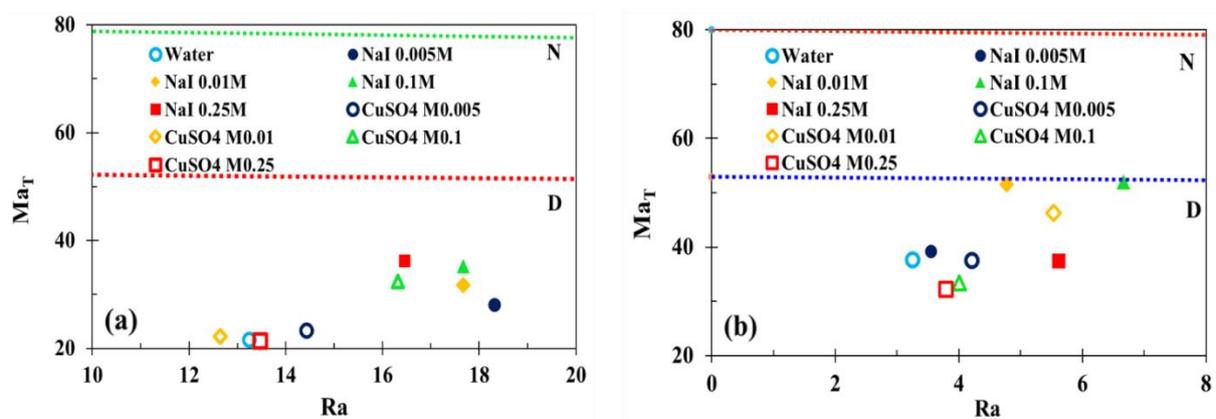

**Figure 9**: Phase plot of the thermal Marangoni and Rayleigh numbers for different droplets on (a) hydrophilic surface and (b) SHS. The lines N and D represent the stability criteria by Nield [38] and Davis [39], respectively.

### 3. e. Behaviour of the internal solutal advection

Figure 10 shows the evolution of salt concentration at the bulk and interface of the droplet with progressing evaporation. The bulk concentration is determined by equating the product of initial concentration and initial volume to the instantaneous concentration and volume. The interfacial concentration is different, as solvated ions preferentially adsorb-desorb to the interface (evident from change of surface tension and contact angle) [47, 48]. The interfacial concentration evolution has been determined following protocol reported by present authors [32]. A clear difference between the concentration of solvated ions at the bulk and interface is notable (figure 10), signifying that solutal advection within the droplet must be present. Further, with progressing evaporation, the concentration difference is enhanced, leading to further accelerated evaporation. This leads to accelerated non-linear shift in the evaporation rate beyond a certain time-frame (refer fig. 3).

The thermal gradient induces advection directed from the rim towards the centre of the droplet (refer fig. 8). The solutal gradients (fig. 10) indicate that the solutal advection will be directed from the centre of the droplet towards the rim. Hence, two opposing advection patterns are theorized. In the hydrophilic case, the thermal component is very weak (fig. 9), the solutal can be theorized to be the dominant advection mode. For the SHS, the thermal advection, though not stable, is significantly stronger (fig. 9), and hence may oppose the solutal counterpart. This could be another plausible reason behind the weak advection behaviour in the concentrated droplets on SHS (which also leads to unhindered nucleation of the salt crystals).

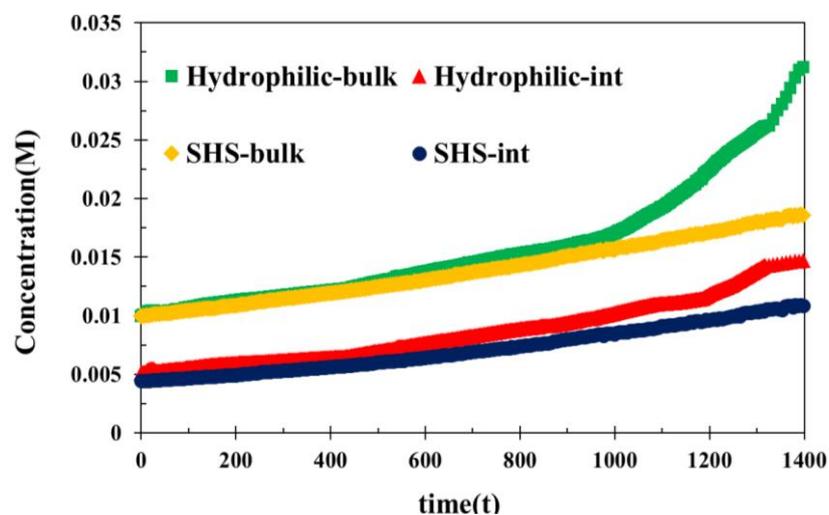

**Figure 10**: Bulk and interfacial (int) concentration of 0.01M NaI solution droplets on hydrophilic surface and superhydrophobic surfaces (SHS) during the evaporation process.

A similar scaling model has been proposed based on the species transport modes inside the droplet to understand the role of solutal Marangoni advection. The species balance equation is expressed as (refer supplementary information for detailed analysis)

$$\rho d_c^2 \dot{\theta} \sec^4(\theta/2) = 16 D \Delta C \cot(\theta/2) \left\{ Ma_s \sec^2(\theta/2) - 4 Ma_s \frac{Sc}{Ca_s} \sin\left(\frac{\theta}{2}\right) \cos\left(\frac{\theta}{2}\right) \right\} \quad (13)$$

Where, D is the diffusion coefficient of the salt in the water, $A_c$ is the contact area of the droplet, $A_s$ is the surface area of the droplet, $\Delta C$ is the concentration difference between the bulk and the interface, $h(t)$ is the height of the droplet ($h(t) = \frac{d_c}{2}\tan(\theta/2)$, refer supplementary information). The $u_c$ denotes the internal circulation velocity due to solutal Marangoni convection and is scaled as [38]

$$u_c = \frac{\sigma_c \Delta C}{\mu} \quad (14)$$

Also difference between the concentration at bulk and that of interface ($\Delta C$) is given as

$$\Delta C = \left(\frac{\rho \mu \dot{h}}{\sigma_c}\right)^{1/2} \text{ for hydrophilic substrate} \quad (15)$$

$$\Delta C = \left(\frac{\rho \mu \dot{d}}{\sigma_c}\right)^{1/2} \text{ for SHS} \quad (16)$$

Where $\dot{h}$ is the change in the droplet height with time in case of hydrophilic substrate droplet, $\dot{d}$ is the time rate of change in the droplet diameter in case of SHS, $\sigma_c$ is the surface tension gradient due to change in the salt concentration (obtained experimentally using pendant drop method for a large number of salt concentrations and then obtaining a correlation).

Further, $Ma_s$ represents the solutal Marangoni number, $Ca_s$ is the solutal Capillary number and $Sc$ is the Schmidt number. These are expressed as

$$Ma_s = \frac{\sigma_c \Delta C d_c}{2 \mu D} \tan(\theta/2) \quad (17)$$

$$Ca_s = \frac{\mu d_c \dot{\theta}}{2\sigma} \sec^2(\theta/2) \quad (18)$$

$$Sc = \frac{\mu}{\rho D} \quad (19)$$

Figure 11 illustrates the map of the $Ma_s$ vs. the $Ma_T$. The criterion of stability of internal advection is reported by Joo [42] and has been represented by two iso-Lewis number lines, the Le=0 and Le-5 (Lewis number is $Le = \alpha/D$, where $\alpha$ is the thermal diffusivity and D is the species diffusion coefficient). All data points lie far to the right away from the low value Le curves, signifying that the mass transport regime is dominant over the heat transport regime for the droplet interior. Further, the points lie to the right of Le=0, signifying stable solutal advection [49]. With increase in salt concentration, the points shift towards the right, and the relative values of the $Ma_s$ is large compared to the $Ma_T$, signifying that the solutal Marangoni advection is the dominant mode, and essentially leads to enhanced evaporation rates. Increase of the $Ma_s$ with salt concentration signifies enhanced internal advection, and is in agreement with PIV results.

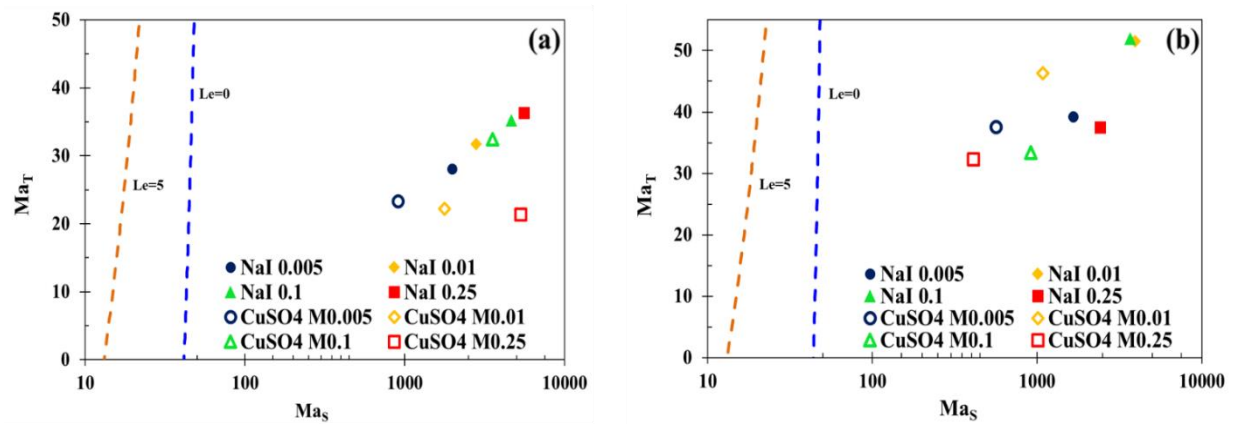

**Figure 11**: Plot of the solutal Marangoni number ($Ma_s$) with thermal Marangoni ($Ma_T$) for different droplets on (a) hydrophilic surface and (b) SHS. The lines represent iso-Le lines [40].

Having established that the solutal Marangoni advection is the genesis of the internal circulation, the spatio-temporally averaged circulation velocities can be mapped from the scaling analysis (from expression of $u_c$). Figure 12 illustrates the comparison between the experimental spatio-temporally averaged internal flow velocities with respect to the theoretical $u_c$ predictions. The velocity values obtained from solutal advection theory are in good agreement with the experimental internal velocities, and further cement the proposition that the solutal advection is the dominant governing mechanism behind the internal circulations and the improved evaporation kinetics.

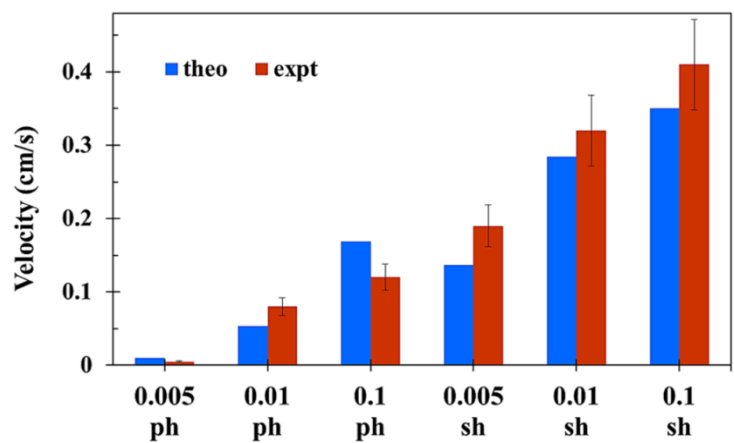

**Figure 12**: Comparison of the experimentally observed internal flow velocities (spatio-temporal mean velocities at droplet mid-plane) and the predicted velocities predicted

(equation (14)) (*ph* indicates hydrophilic surface and *sh* indicates SHS) for NaI solution droplets.

**3. f. Scaling the interfacial shear modified Stefan flow**

The classical vapour-diffusion models were found to be incapable to quantify the augmented evaporation rates of the saline droplets. It is proposed that the shear induced within the vapour layer due to internal advection leads to improvement of the Stefan flow [50] from the evaporating droplet, leading to augmented mass transfer. A semi-analytical approach based on the Stefan flow surrounding the droplet, and its modification due to the shear at the interface due to internal advection, has been proposed to determine the evaporation rates. The amount of liquid evaporated from the interface (liquid side) enters the vapour phase shrouding the droplet. From mass conservation across the droplet interface (liquid side to gas side), the following holds true

$$\dot{m}_L = \rho_L \dot{V} = \dot{m}_g = \rho_g A_s u_g \tag{20}$$

where $\dot{m}_L$ is the mass rate of the evaporating liquid, $\dot{m}_g$ is the mass rate of the vapour escaping across the interface, $\rho_L$ is the density of the liquid, $\rho_g$ is the density of the vapour, $A_s$ is the surface area of the droplet, $u_g$ is the Stefan flow velocity of the vapour in the gaseous phase, $\dot{V}$ is the rate of liquid volume evaporating away.

Eqn. 20 is a representation of the Stefan flow around a droplet evaporating in quiescent media [43]. The balance of shear stresses across the liquid-vapour interface yields

$$\mu_L \frac{\partial u}{\partial y}\bigg|_L = \mu_g \frac{\partial u}{\partial y}\bigg|_g \tag{21}$$

Where $\mu_L$ is the dynamic viscosity of the liquid, $\mu_g$ is the dynamic viscosity of the vapour, and $\frac{\partial u}{\partial y}$ is the shear rate (where *l* and *g* represent the liquid and gas sides). The internal advection velocity inside evaporating water droplets is low, ~0.03-0.04 cm/s (in agreement with literature [45]). Hence, the shear driven velocity of vapour in the gas phase can be obtained from eqn. 21. However, since the thickness of the vapour layer surrounding the droplet is not easily determined, the eqn. 21 requires to be scaled. Let $u'$ represent the average velocity of the effective Stefan flow (Stefan flow and the interfacial shear induced flow) surrounding the saline droplet. From eqn. 21, the scaling yields

$$\mu_L \frac{u_d'}{h'} = \mu_g \frac{u'}{x} \tag{22}$$

Where $u_d'$ is the average internal velocity of advection within saline droplet, $h'$ is the droplet height and $x$ is some distance in the vapour diffusion layer above the droplet surface (essentially a characteristic length).

Now, mass conservation at the interface of the saline droplet also yields

$$\dot{m}' = \rho_L \dot{V}' = \dot{m}_g' = \rho_g A'_s u' \tag{23}$$

Where $\dot{m}'_g$ is the mass rate of the vapour entering in the gaseous phase from saline droplet, $A'_s$ is surface area of the saline droplet, and $\dot{V}'$ is the volume evaporation rate of the saline droplet. Assuming that the density of the saline solution is same as water, and the thickness of the vapour shroud is unchanged, eqns. 20, 21, 22 and 23 can be used to eliminate x, leading to

$$\dot{V}' = \left(\frac{A'_s h' u_d'}{A_s h_w u_w}\right) \dot{V} \qquad (24)$$

Where, $u_w$ and $h'$ represent the average advection velocity (solutal advection) within evaporating water droplet, and height of the water droplet. Figure 13 compares the experimentally observed volume evaporation rate with the scaled values (eqn. 22) and good prediction accuracies are noted for both wetting states. The fact that the solutal advection velocity also predicts the improved evaporation rate to appreciable extents further establishes its role as the dominant mechanism behind the internal advection.

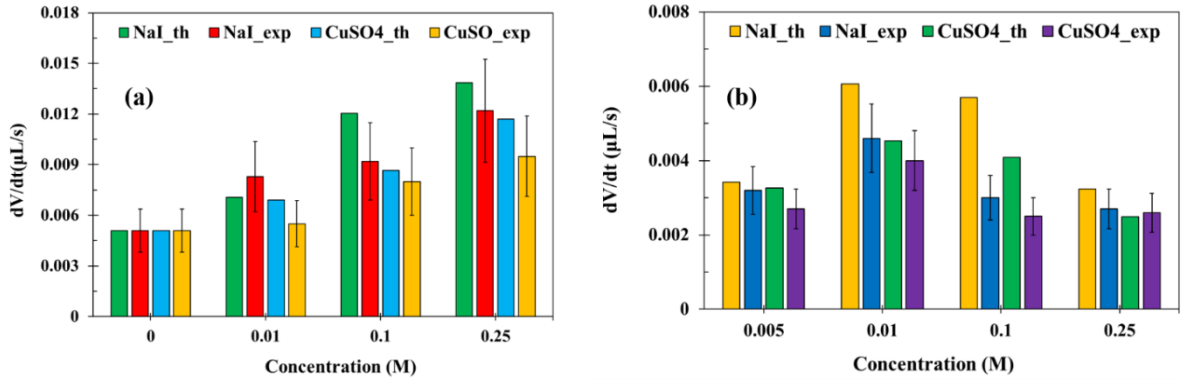

**Figure 13**: Comparison of the experimentally evaporation rate with scaled Stefan flow based model predictions for NaI and CuSO$_4$ solution droplets on (a) hydrophilic surface and (b) SHS.

## 4. Conclusions

The present article reports detailed experimental observations and theoretical analysis on the evaporation kinetics of saline sessile droplets on hydrophilic and SH surfaces. Aqueous NaI and CuSO$_4$ solutions were used as test fluids (based on previous reports by the authors. On hydrophilic substrates the evaporation rate increases as a direct function of the salt concentration in the droplet, while on SHS, the evaporation rate initially increases with the concentration and then decreases. The classical vapour-diffusion driven model for sessile droplets was noted incapable to explain the modified evaporation rates. The models consider the surface tension and contact angle change due to presence of salt, and yet were not able to predict the evaporation rates. PIV study shows that the interior of the droplet exhibits advection behaviour, which improves with salt concentration. Droplets on both the wetting states exhibit the typical internal flow behaviour, and the possible reasons are cited as thermal or solutal Marangoni circulation. Further, counter-intuitive arrest of advection is noted for high concentration droplets on SHS, and inception of crystal nucleation on the SHS has been shown as the plausible reason.

The genesis of the internal advection has been probed through scaling analysis. The energy and species transport modes have been scaled to form a mathematical framework for the thermal and solutal Marangoni advection. Experimental determination of the thermal and solutal gradients within the droplets has also been performed. Stability maps show that the thermal Rayleigh advection is not a plausible mechanism, and the thermal Marangoni advection is also a weak mechanism to generate the noted internal flows. The analysis yields that the solutal Marangoni advection is the dominant cause for the flows within. This is further proven as the solutal model is able to predict the internal flow velocities with good accuracy. Additionally, an interfacial shear driven modified Stefan flow based mathematical formalism is proposed to predict the modified evaporation rates by scaling the observed internal circulation strength. The deduced theoretical values match well with the experimentally observed evaporation rate. This further confirms the dominant role of the internal advection on the modified evaporation behaviour. The findings could have strong implications in droplet based macro and microfluidic systems and devices.

**Supplementary Material:** The associated supplementary material document contains the detailed mathematical description of the droplet evaporation rates, the thermal imaging of the evaporating droplets, discussions on microscopy studies on the droplets, etc.

**Competing interests:** We declare we have no competing interests with respect to this research work.